
\def\today{\ifcase\month\or January\or February\or March\or
April\or May\or June\or July\or August\or September\or
October\or November\or December\fi \space\number\day,
\number\year}



\font \bbrm=cmbx10 at 12pt

\def\bigtype{\bbrm}

\hsize=13.5cm
\magnification=1200
\def\ce{\centerline}

\def\title #1{\null\bigskip\ce{\bigtype #1}\bigskip}

\def\alp{\alpha}                
\def\bet{\beta}
\def\gam{\gamma}                
\def\del{\delta}

\def\lam{\lambda}               \def\Lam{\Lambda}
                \def\Sig{\Sigma}

\def\ome{\omega}                


\def\calE{{\cal E}}
\def\calF{{\cal F}}

\def\calU{{\cal U}}




\font\tenboldgreek=cmmib10  \font\sevenboldgreek=cmmib10 at
7pt
\font\fiveboldgreek=cmmib10 at 7pt
\newfam\bgfam
\textfont\bgfam=\tenboldgreek \scriptfont\bgfam=\sevenboldgreek
\scriptscriptfont\bgfam=\fiveboldgreek


\font\tengerman=eufm10 \font\sevengerman=eufm7 \font\fivegerman=eufm5
\font\tendouble=msbm10 \font\sevendouble=msbm7 \font\fivedouble=msbm5

\textfont4=\tengerman \scriptfont4=\sevengerman
\scriptscriptfont4=\fivegerman
\newfam\dbfam
\textfont\dbfam=\tendouble \scriptfont\dbfam=\sevendouble
\scriptscriptfont\dbfam=\fivedouble
\def\gr{\fam4}

\mathchardef\ng="702D
\mathchardef\dbA="7041
\mathchardef\sm="7072
\mathchardef\nvdash="7030
\mathchardef\nldash="7031
\mathchardef\lne="7008
\mathchardef\sneq="7024
\mathchardef\spneq="7025
\mathchardef\sne="7028
\mathchardef\spne="7029
\mathchardef\ltms="706E
\mathchardef\tmsl="706F

\mathchardef\dbA="7041


       \def\grg{{\gr g}}

\mathchardef\dbA="7041 
\mathchardef\dbB="7042 
\mathchardef\dbC="7043 
\mathchardef\dbD="7044 
\mathchardef\dbE="7045 
\mathchardef\dbF="7046 
\mathchardef\dbG="7047 
\mathchardef\dbH="7048 
\mathchardef\dbI="7049 
\mathchardef\dbJ="704A 
\mathchardef\dbK="704B 
\mathchardef\dbL="704C 
\mathchardef\dbM="704D 
\mathchardef\dbN="704E 
\mathchardef\dbO="704F 
\mathchardef\dbP="7050 
\mathchardef\dbQ="7051 
\mathchardef\dbR="7052 
\mathchardef\dbS="7053 
\mathchardef\dbT="7054 
\mathchardef\dbU="7055 
\mathchardef\dbV="7056 
\mathchardef\dbW="7057 
\mathchardef\dbX="7058 
\mathchardef\dbY="7059 
\mathchardef\dbZ="705A \def\ZZ{{\fam=\dbfam\dbZ}}

\def\nek{,\ldots,}
\def\sdp{\times \hskip -0.3em {\raise 0.3ex
\hbox{$\scriptscriptstyle |$}}} 


\def\Hom{\mathop{\rm Hom}\nolimits}









\def\tilf{{\widetilde f}}

\def\tilK{{\widetilde K}}

\def\tilP{{\widetilde P}}

\def\tilU{{\widetilde U}}


\def\tilvarphi{{\widetilde\varphi}}

\def\sqr#1#2{{\vcenter{\hrule height.#2pt\hbox{\vrule
width.#2pt height#1pt \kern#1pt \vrule width.#2pt}\hrule
height.#2pt}}}

\def\buildrul#1\under#2{\mathrel{\mathop{\null#2}\limits_{#1}}}

\def\boxit#1{\vbox{\hrule\hbox{\vrule\kern3pt\vbox{\kern3pt#1
\kern3pt}\kern3pt\vrule}\hrule}}

\def\subheading#1{\medskip\goodbreak\noindent{\bf
#1.}\quad}
\def\sect#1{\goodbreak\bigskip\centerline{\bf#1}\medskip}

\def\longmapright #1 #2 {\smash{\mathop{\hbox to
#1pt {\rightarrowfill}}\limits^{#2}}}
\def\longmapleft #1 #2 {\smash{\mathop{\hbox to
#1pt {\leftarrowfill}}\limits^{#2}}}

\def\ref#1{\par\smallskip\hang\indent\llap{\hbox
to \parindent{#1\hfil\enspace}}\ignorespaces}

\def\back{{\raise 2.5pt\hbox{$\,\scriptscriptstyle\backslash\,$}}}

\def\part{\partial}
\def\lwr #1{\lower 5pt\hbox{$#1$}\hskip -3pt}
\def\rse #1{\hskip -3pt\raise 5pt\hbox{$#1$}}
\def\lwrs #1{\lower 4pt\hbox{$\scriptstyle #1$}\hskip -2pt}
\def\rses #1{\hskip -2pt\raise 3pt\hbox{$\scriptstyle #1$}}

\def\<#1{\left\langle{#1}\right\rangle}

\def\subinbn{{\subset\hskip-8pt\raise
0.95pt\hbox{$\scriptscriptstyle\subset$}}}

\def\llvdash{\mathop{\|\hskip-2pt \raise 3pt\hbox{\vrule
height 0.25pt width 1.5cm}}}

\def\lvdash{\mathop{|\hskip-2pt \raise 3pt\hbox{\vrule
height 0.25pt width 1.5cm}}}

\def\fakebold#1{\leavevmode\setbox0=\hbox{#1}%
  \kern-.025em\copy0 \kern-\wd0
  \kern .025em\copy0 \kern-\wd0
  \kern-.025em\raise.0333em\box0 }

\font\msxmten=msxm10
\font\msxmseven=msxm7
\font\msxmfive=msxm5
\newfam\myfam
\textfont\myfam=\msxmten
\scriptfont\myfam=\msxmseven
\scriptscriptfont\myfam=\msxmfive
\mathchardef\rhookupone="7016
\mathchardef\ldh="700D
\mathchardef\leg="7053
\mathchardef\ANG="705E
\mathchardef\lcu="7070
\mathchardef\rcu="7071
\mathchardef\leseq="7035
\mathchardef\qeeg="703D
\mathchardef\qeel="7036
\mathchardef\blackbox="7004
\mathchardef\bbx="7003
\mathchardef\simsucc="7025

\def\bigsquare{{\fam=\myfam\bbx}}

\font\tencaps=cmcsc10
\def\smallcaps{\tencaps}

\def\author#1{\bigskip\ce{\smallcaps #1}\medskip}

\null
\overfullrule=0pt
\sect{POINCAR\'E--BIRKHOFF--WITT THEOREM}
\vskip-0.50truecm
\sect{FOR QUADRATIC ALGEBRAS OF KOSZUL TYPE}
\ce{A.~Braverman and D.~Gaitsgory}
\bigskip
\ce{School of Mathematical Sciences}
\ce{Tel Aviv University}
\ce{Ramat Aviv,  69978, Israel}
\bigskip

\sect{0. Introduction}

\subheading{0.1. Homogeneous quadratic algebras} Let $ V $ be a vector space
 over some field $ k $ and let $ T(V) ={}\oplus T^i $ be its tensor algebra
over
$k$. Fix a subspace $R \subset T^2 = V\otimes V$, consider the two-sided ideal
$J(R)$ in $T(V)$ generated by $R$ and denote by $Q(V,R)$ the quotient algebra
$T(V)/J(R)$. This is what is known as (a homogeneous) {\it quadratic algebra}.

\subheading{0.2. Nonhomogeneous quadratic algebras} In a similar way we define
{\it nonhomogeneous quadratic algebras}, the main objects of our study. They
are  {\it filtered} analogs of graded homogeneous quadratic algebras.

Consider the
natural filtration $F^i(T) = \{\oplus T^j|j \leq i \}$ of $T(V)$. Fix a
subspace
$P\subset F^2(T) = k \oplus V \oplus (V\otimes V) $, consider the two-sided
ideal
$J(P)$ in $T(V)$ generated by $P$ and denote by $Q(V,P)$ the quotient algebra
$T(V)/J(P)$. The algebra $Q(V,P)$ will be called a {\it nonhomogeneous
quadratic
algebra}.

\subheading{0.3} Let $U = Q(V,P)$ be a nonhomogeneous quadratic algebra. It
 inherits a filtration $U_0 \subset U_1 \subset \dots \subset U_n
\dots$ from $T(V)$. We would like to describe the associated graded algebra
gr$\, U$.

Consider the natural projection $p:F^2(T) \to V\otimes V $ on the homogeneous
component, set $R = p(P)$ and consider the homogeneous quadratic algebra $A =
Q(V,R)$.  We have the natural epimorphism $p:\ A \to {\rm gr}\; U $.

\subheading{Definition}
 A nonhomogeneous quadratic algebra $U = Q(V,P)$,
or, more precisely, the subspace $P$ of $F^2(T) $, is of
{\it Poincar\'e--Birkhoff--Witt (PBW) type} if the natural
projection $p:\ A = Q(V,R) \to$ gr $U$ is an isomorphism.

\subheading{0.4. Lemma}
{\sl Suppose that the subspace $P \subset F^2(T)$ is of
PBW-type. Then it satisfies the following conditions :

(I) $ P \cap F^1(T) = 0$;

(J) $ (F^1(T)\cdot P \cdot  F^1(T)) \cap   F^2(T) = P$}.

\noindent Proof is straightforward.

\subheading{0.5. The main theorem}
{\sl Let $U=Q(V,P)$ be a nonhomogeneous quadratic
algebra. Take $R=p(P) \subset T^2(V)$ and consider
the corresponding homogeneous quadratic algebra $A=Q(V,R)$.

Suppose $A$ is a Koszul algebra
(see 3.4). Then conditions (I) and (J) above imply that
the subspace $P$ and hence the algebra $U$, is of PBW-type.}
\smallskip

The theorem will be proved in section 4 by means of
deformation theory; in paticular, it will provide a new
proof of the classical PBW-theorem. Note, that a slightly weaker version of the
above result (for $P \subset T^2(V)\oplus T^1(V)$) was proved by
Polishchuk and Posetselsky ([PoP]) by methods different from ours.

\subheading{0.6. Example}
If $A=S(V)$  (the symmetric algebra of $V$) and $P$ does not
have the scalar component, then $R$ is equal to $\Lam ^2(V)$,
i.e.,  the space of all antisymmetric 2-tensors (more precisely,
it is the space generated by all tensors of the form
$x\otimes y-y\otimes x$; this is important when char $\/k=2$).
In this case $P$ can be represented as a graph of some map
$\alp:\Lam ^2V\to V$. The classical Poincar\'e--Birkhoff--Witt theorem asserts
then that $P$ is of PBW-type if and only if $\alp$ satisfies the Jacobi
identity.
In \S  4 we will see that this is equivalent to condition (J) (which explains
this abbreviation).

\subheading{0.7. Contents} This paper is organized as follows: in \S \S  1 and
2
we briefly explain the idea of our proof and review some basic facts concerning
deformations and cohomology of associative algebras. In \S  3
we define Koszul algebras and reformulate the main problem in terms
of certain explicit identities (which can be considered as generalised
Jacobi identities) and in \S  4 we prove the above theorem. Since
we use a somewhat nontraditional definition of Koszul algebras we
prove its equivalence with the usual one in the appendix together
with a review of some other basic properties of Koszul algebras.

\subheading{0.8. Acknowledgments}
It is a great pleasure for us to express our deep gratitude to
Joseph Bernstein for posing the problem and valuable discussions.
We are also grateful to Arkadii Vaintrob who introduced us to Koszul algebras.

\sect{1. PBW-theorem and Deformations}

\subheading{1.1} First, we will briefly explain the idea of the proof
of Main Theorem.
We start with the following simple observation.  Suppose we
are given a $k$-algebra $U$  with an increasing filtration:
$U=\bigcup\limits^\infty_{i=0}U_i$  and let gr $ U$ be its
associated graded algebra:
gr $ U=\bigoplus\limits^\infty_{i=0}U_i/U_{i-1}$.
Then we can always construct a family of  associative
filtered algebras over the affine line  Spec($k[t]$)
(this family is an algebra over the ring $k[t]$)
such that {\it its fiber over point $t=0$ is isomorphic to gr$ U$ and
its fiber over any point $t=\lam\ne 0$ is
isomorphic to $U$}.
Namely, we define this algebra ${\cal U}$ as follows:
$$
{\cal U}=\{ \Sig u_i t^i\mid u_i\in U_i\}\subset U\otimes_kk[t]=U[t].
$$
(The algebra $\calU$ is usually called the Rees ring of $U$). The verification
of the property italicized is
straightforward. One can immediately see that $\calU$ becomes a graded
$k[t]$-algebra if we set deg$\, t=1$.
(It is easy to see that in the situation described
above every fiber over $t\neq 0$ has a natural structure of a filtered algebra
and {\it not} of a graded one since deg$\, t=1$.)

The idea of the proof of the theorem is to reverse this
process -- namely, starting with a quadratic algebra $A=Q(V,R)$,
where $R\subset T^2(V)$ and $P$ are as in the introduction, to construct
a family of filtered algebras over Spec($k[t]$) such that its
fiber at $t=0$ is $Q(V,R)$, and its fiber at $t=1$ is $Q(V,P)$
and such that the corresponding associated graded family is trivial.
To do this rigorously we need the notion of a {\it graded deformation}.

\subheading{1.2. Graded deformations}
Let $A$ be a (positively) graded associative algebra over a field $k$.
By an {\it $i$-th level graded deformation} of $A$
we will mean a graded $k[t]/k[t]t^{i+1}$-algebra $A_i$, which is free
as a $k[t]/k[t]t^{i+1}$-module, together with an isomorhism
of $A_i/tA_i\cong A$. (Here deg $t=1$.)
By a {\it graded deformation} $A_t$ of $A$ we mean a graded algebra over the
polynomial ring $k[t]$ (remember that deg $t=1$), which is free as a module
over this ring, together with an isomorphism $A_t/tA_t\cong A$.

\subheading{1.3}
Let $\calE(A)$ denote the category
of all graded deformations of $A$ where the morphisms are by definition
isomorphisms of graded $k[t]$-algebras (by definition this category is a
groupoid). Analogously let $\calE_i(A)$
denote the {\it groupoid of all $i$-th level graded deformations of $A$}.
 We denote
by $\calF_i$ the functor from $\calE_{i+1}(A)$ to $\calE_i(A)$ given by
reduction
modulo $t^i$.
The following lemma is staightforward.
\subheading{Lemma} {\sl Reductions modulo  $t^i$ define an equivalence between
the category
$\calE(A)$ and the inverse limit of the categories $\calE_i(A)$
with respect to the functors $\calF_i$.}



\subheading{1.4}
Graded deformations have the following property which explains
their importance for us: for every $\lam\in k$ the fiber of ${\cal A}$
at $\lam$ has a natural structure
of a filtered algebra and its associated graded algebra gr$(A_\lam)$ is
canonically isomorphic to the original algebra $A$.

How to continue graded deformations from one level to another? The following
proposition answers the question.

\subheading{1.5. Proposition}
a){\sl The set of isomorphism classes of objects of $\calE_1$ canonically
identifies with $H^2_{-1}(A,A)$.}

(Here $H^i(A,A)$ denotes the graded Hochschild cohomology of $A$ with
coefficients in $A$ (see 2.1) and subscript denotes the
natural grading induced on these cohomology groups by the grading of $A$.)

b) {\sl Let $A_i$ be an object of $\calE_i(A)$. Then the obstruction for its
continuation to the
$(i+1)$-st level lies in $H^3_{-i-1}(A,A)$.}

c) {\sl Let $A_i$ be as in (b). Then the set of isomorphism
classes of continuations of $A_i$ to the $(i+1)$-st level has a natural
structure of an
$H^2_{-i-1}(A,A)$-homogeneous space.}

\noindent Proof of this proposition will be sketched in the next section.

\subheading{1.6. Remark} All definitions and statements of 1.2--1.5 are
essentially contained in the classical work [G] of M.~Gerstanhaber. Here we
need
a slightly different version of [G] relevant for the graded situation.

\sect{2. Cohomology and Deformations}

Here we review some facts concerning cohomology of
associative algebras and prove the proposition stated in the previous section.

\subheading{2.1 Hochschild cohomology} Let $A$ be any associative $k$-algebra
 and $M$  any $A$-bimodule,
i.e., , left $A\otimes A^\circ$-module (where $A^\circ$ is the
same algebra $A$ with multiplication $a\cdot b= ba$).  We will be
interested in $Ext_{A\otimes A^\circ}(A,M)$. This cohomology
can be computed using the resolution $B^i(A)$ of $A$ ({\it bar
resolution})
$$
\matrix{
B^i(A)=A\otimes T^i(A)\otimes A\quad\left(T^i(A)=A^{\otimes
i}\right)\cr
\part^i\left( a_0\otimes a_1\cdots \otimes a_i\otimes
a_{i+1}\right)=\sum^i_{k=0}(-1)^k(a_0\otimes a_1\otimes\cdots\otimes
a_ka_{k+1}\otimes\cdots\otimes a_{i+1})\ .
}
$$
One can easily verify that it gives a projective
(in fact, even free) resolution of $A$  as an
$A\otimes A^\circ$-module ($B^i(A)$
has natural structure of a free left $A\otimes A^0$-module).
It is obvious that
$\Hom_{A\otimes A^0}\left(B^i(A),M\right)$  is
the same as $\Hom_k\left(T^i(A),M\right)$
for any $A$-bimodule $M$.
Hence, we see that $Ext_{A\otimes A^\circ}(A,M)$ can be
computed by means of the complex:
$C^i(A,M)=\Hom\left(T^i(A),M\right)$ with the differential
$$
\matrix{df(a_1,\cdots ,a_{i+1})=a_1f\left(a_2,\cdots ,a_{i+1}\right)+
(-1)^{i+1}f(a_1\nek a_i)a_{i+1}+\cr
\sum^i_{k=1}(-1)^kf\left(a_1, \cdots a_ka_{k+1}\nek a_{n+1}\right),\;  {\rm
for}\; f\in C^i(A,M). }
$$
This cohomology is called the {\it Hochschild cohomology} of $A$  with
coefficients in $M$.  We denote it by $H^i(A,M)$.

\subheading{2.2. Graded Hochschild cohomology}Let now $A$ be a graded algebra
and let $M$ be a graded $A$-bimodule.
In this case we shall slightly modify the above definition of $H^i(M)$.
Namely, we set $H^i(M)$ to be equal to $Ext_{A\otimes A}(A,M)$ in the category
of graded $A$-bimodules. This cohomology can be computed by means of the
complex
$C^{\cdot}_{gr}(A,M)$ (subscript "gr" means "graded") where
$$C^i_{gr}=\bigoplus _{j=-\infty}^{\infty}Hom_j(T^i(A),M)
$$
and Hom$_j(T^i(A),M)$ is the set of all homogeneous maps of degree of
homogenuity
$j$ (the differential is defined as in 2.1). From now on (except for subsecton
5.4) we shall deal only
with graded algebras and modules, so, "Hochschild cohomology" will
always mean "graded Hochschild cohomology".
\subheading{2.3. Proof of Proposition 1.5(a)}
When we identify $A_1$ (our first level deroramation) with $A\oplus A\cdot t$,
the
multiplication in $A$ is described by a $k$-linear map $f$: $A\otimes A\to A$,
 namely
$$
a\times b=ab+tf(a,b)\quad ({\rm deg}\; (f)=-1).
$$
The associativity condition is rewritten in terms of $f$ as:
$$f(a,b)c-f(ab,c)+f(a,bc)-af(b,c)=0\; {\rm for\ any}\; a, b, c\in A.
$$
In other words, $f$ must be a Hochschild 2-cocycle. It is easy to check that
two infinitesimal deformations are isomorphic if and only if the corresponding
cocycles are cohomologous.

\subheading{2.4. Proof of Proposition 1.5(b)} Suppose we have an $n$-th level
deformation $A_n$. If we identify $A_n$ with $A\otimes {k[t]/k[t]\/t^{n+1}}$,
then the multiplication in $A_n$ is described in terms of a sequence of maps
$f_i$: $A\otimes A \to A$, deg ($f_i$)=$-i$, $i=1,...,n$ (the product of any
two
elements $a$ and $b$ of $A$ in $A_i$ is defined by
$$
a\times b=ab+\sum_{i=1}^n f_i(a,b)t^i.
$$
If $f_{n+1}$ defines a continuation of $A_n$, then it satisfies:
$$\sum^n_{i=1}
f_i(f_{n-i}(a,b),c)-f_i(a,f_{n-i}
(b,c))=f_{n+1}(a,bc)-f_{n+1}(ab,c)+f_{n+1}(a,b)c-af_{n+1}(b,c)
$$
for any $a, b, c\in A$. In other words, $f_{n+1}$ must have a prescribed
coboundary, namely the left hand side of the above formula, whose image in
cohomology is precisely the obstruction to the existence of a continuation from
the $i$-th level to the next one (it is not difficult to show that the above
expression is always a cocycle).

\subheading{2.5. Proof of Proposition 1.5(c)} As in (b) we see, that we may
vary
 $f_{n+1}$ by a cocycle. It
is also easy to verify that two continuations are isomorphic if
the corresponding $f_{n+1}$'s are cohomologous. This means that
the second cohomology group acts transitevely on the set of
isomorphism classes of continuations.

\sect{3. Koszul Algebras and Certain Identities}

\subheading{3.1} Now let us return to quadratic algebras.
Let $V$ be a vector space over $k$ and let $P\subset V$
satisfy the conditions (I) and (J) of 0.2. Let $R$ be the projection of $P$ to
the second
homogeneous component. Now we ask: is the natural surjection from $Q(V,R)$ to
gr$(Q(V,P))$ an isomorphism? In other words, under what conditions is $P$ of
PBW
type?

In this section  we shall reformulate this problem and define the class of
algebras
(Koszul algebras) for which we can completely answer the above question.

\subheading{3.2} As it was said in the introduction, conditions (I) and (J) are
 necessary for $P$ to be of PBW-type.
In fact, they only require the injectivity of the map $p$: $Q(V,R)\to grQ(V,P)$
on the 3-rd term of the natural increasing filtration arising from the grading.
{}From now on we will assume that all nonhomogeneous quadratic algebras we
consider satisfy (I). Then the subspace $P\subset F^2(V)$ can be described in
terms of two maps $\alp$: $R\to V$ and $\bet$: $R\to k$ as $P= \{x-\alp
(x)-\bet
(x)\mid x\in R \}.$

\subheading{3.3. Lemma} {\sl The condition (J) can be rewritten in terms of
$\alp$ and $\bet$ in the following way:

\item{(i)} $Im(\alp\otimes id-id\otimes \alp )\subset R$
(this map is defined on $(R\otimes V)\cap (V\otimes
R)$);
\item{(ii)} $\alp\circ (\alp\otimes id -id\otimes\alp) =-(\bet\otimes
id -id\otimes \bet)$;
\item{(iii)} $\bet\circ (id\otimes \alp -\alp\otimes id)\equiv 0$}.

\noindent {\bf Proof:}\quad  Let $x\in (R\otimes V\cap V\otimes R)$. Then by
definition of $\alp$ and $\bet$ we have:
$$
\eqalign{&\alp\otimes id(x)+
\bet\otimes id(x)-x\in P\cdot T_1;\cr
&id\otimes\alp (x)+id\otimes\bet (x)-x\in T_1\cdot P.\cr}
$$
Thus by (J) we have
$$
\alp\otimes id(x)-id\otimes \alp(x)+
\bet\otimes id(x)-id\otimes \bet (x)\in P.
$$
This implies that
$$
\eqalign{&\alp\otimes id(x)-id(x)\otimes\alp (x)\in R;\cr
&\alp (\alp\otimes id(x)-id\otimes\alp (x))=-(\bet\otimes id(x)-
id\otimes\bet (x);\cr
&\bet(\alp\otimes id(x)-id\otimes\alp (x))=0\quad \bigsquare}
$$

Our main theorem asserts, that conditions (i),
(ii) and (iii) above
enable us to build a graded deformation of $Q(V,R)$ such that its fiber
at the point $t=1$ is canonically isomorphic, as a filtered algebra,
to $Q(V,P)$, provided $Q(V,R)$ is a Koszul algebra. In view of Remark
1.4 this means that our $Q(V,P)$ is of PBW type.

\subheading{3.4. Definition}
Let $A=Q(V,R)$ be a quadratic algebra. We say that it is of
{\it Koszul type} if $H^i_j(A,M)=0$ for all $i<-j$ and for every $\ZZ
^+$-graded
$A$-bimodule $M$.

\subheading{3.5. Example} Let $V$ be a vector space. Denote by $S(V)$
its symmetric algebra  with its usual grading. Then it is well known that
$H^i(S(V),M)$ is a subquotient of
 $M\otimes \Lam^i(V^*)$, where the elements of $\Lam^i(V^*)$ have degree $-i$.
Thus, one can see that $H^i_j(S(V),M)$ vanishes for $i<-j$, so the symmetric
algebra is a Koszul algebra.

\subheading{3.6. Koszul complex} There are several definitions of
Koszul algebras. Some of them and a proof of their equivalence are discussed in
the appendix. Here we shall also need the following one.

Let $A=Q(V,R)$. Define the following subcomplex $\tilK ^{\cdot}(A)$ of
$B^{\cdot}(A)$. Set
$$
K^i(A)=\bigcap^{i-2}_{j=0}V^{\otimes j}\otimes R\otimes
V^{\otimes i-j-2};\quad  \tilK^i(A)=A\otimes K^i\otimes A.
$$
We have a natural imbedding of $K^i(A)$ into $T^i(V)$; hence, into
$T^i(A)$. Thus, we can regard $\tilK^i$  as an $A\otimes A^\circ$-submodule
of $B^i(A)$.  One can verify by a direct computation that
$\tilK^\cdot(A)$  forms a subcomplex of $B^\cdot(A)$. It is called the
{\it Koszul complex associat}ed to $A$. This complex is not  always acyclic.
However, the following is true:

\subheading{Claim}{\sl $A$ is a Koszul algebra if and only if
$$
H^i\left(\tilK^\cdot(A)\right)=\cases{0&if $i>0$\cr
A&if $i=0$\cr}
$$
In other words, $A$ is Koszul iff the map of complexes
$\tilK^\cdot(A)\to B^\cdot(A)$ is a quasiisomorphism.}

\medskip
This claim will be proven in the appendix.

\subheading{3.7. Deformations of Koszul algebras} In the proof of the main
theorem we shall need the following statement about functors $\calF_i(A)$
for a Koszul algebra $A$.
\subheading{Proposition}{\sl Let $A$ be a Koszul algebra. Then

(i) the functors $\calF _i(A):\, \calE_{i}(A)\to \calE_{i-1}(A)$ are
monomorphic on isomorphism classes of objects
for $i>2$

(ii) the functors $\calF _i(A)$ are epimorphic on isomorphism classes of
objects
for $i>3$}
\medskip

\noindent {\bf Proof:} It follows from the definition of Koszul algebras that
$H^2_{-i-1}(A,A)$ vanishes for $i>2$ . Hence Proposition 1.5(c) implies
(i). Analogously, $H^3_{-i-1}(A,A)$ vanishes for $i>3$, hence Proposition
1.5(b)
implies (ii). $\bigsquare$

\subheading{3.8. Remark} In fact one can prove that under the conditions of
3.7. the functors $\calF_i(A)$ are also surjective on morphisms for $i>3$ and
hence, the categories $\calE(A)$ and $\calE_i(A)$ have the same isomorphisms
classes of objects. The reason for that is again cohomologous and it will not
be explained in this paper.

\sect{4. The Main Theorem}
\subheading{4.1. Theorem}
{\sl Let $A=Q(V,P)$ be a Koszul algebra  and let $\alp :R\to V$ and $\bet
:R\to k$
be linear maps that satisfy identities (i), (ii) and (iii) of 3.3.
Then there exists a graded deformation of $A$ such that its fiber at the point
$t=1$ is canonocally isomorphic to $U=Q(V,P)$ defined by means of of $\alp$
and $\bet$.}

\subheading{4.2. Plan of the proof}First we show in subsection 4.3 that a map
$\alp$ satisfying 3.3(i) defines a first level graded deformation of $A$
uniquely up to an isomorphism. Next we show that a map $\bet$, satisfying
3.3(ii) defines a continuation of this deformation to the second level (again,
uniquely up to an isomorphism).
Then condition 3.3(iii) ensures that this deformation lies in the image of the
functor $\calF_3$, i.e. it can be continued to the third level. Thus
Proposition 3.7 implies that it extends to some graded deformation
of $A$, which in fact in view of Remark 3.8 will be uniquely defined. Finally
we prove  that this deformation satisfies the conditions of 4.1.
\subheading{4.3. Extension to the first level}
Both $\alp$ and $\bet$ are defined on $R$ which is equal to $K^2$.  Obviously,
we may regard them as $A\otimes A^\circ$-linear maps from
$\tilK^2$  to $A$.  We denote these maps also by $\alp$ and
$\bet$.  Then we see that condition (i) means that $d\alp=0$
(i.e. $\alp$ is a $2$-cocycle in the complex $\Hom_{A\otimes
A^\circ}(\tilK^\cdot,A)$).  Thus, it defines a cohomology class in
$H^2(A,A)$.

Let us prove that we can find a Hochschild $2$-cocyle
$f_1$  such that $f_1|_{K^2}=\alp$ and $f_1$  is homogeneous
(of degree $-1$). Let $\tilf _1$ be any Hochschild 2-cocycle
which defines the same cohomology class as $\alp$. Then
$\tilf _1-\alp |_{K^2}=d\ome$ for some Koszul 1-cochain $\ome$
of degree $-1$. Next,  we can find a Hochschild 1-cochain $\ome '$
of degree $-1$ such that $\ome '|_{K^1}=\ome$. Define $f_1$
to be $\tilf _1-d\ome '$. We set this $f_1$  to be the first term of
the deformation.
\subheading{4.4. Extension to the second level}
Now, by (ii) $d\bet=\alp\circ (\alp\otimes id-id\otimes \alp)$ in
the Koszul complex.
Let us prove that there exists a Hochschild 2-cochain $f_2$
of degree $-2$ satisfying
$$
df_2=f_1\circ (f_1\otimes id-id\otimes f_1)\quad {\rm and}\quad f_2|_{K^2}=\bet
{}.
$$
We can find a Hochschild $2$-cohain $\tilf _2$ such
that $\tilf _2|_{K^2}=\bet$.
Consider the cocycle $\gam =d\tilf _2-f_1\circ(f_1\otimes id-id\otimes f_1)$.
Then it follows from (ii) that $\gam$ is cohomologically trivial
since $\gam |_{K^3}=0$. Hence, there exists a Hochschild 2-cochain
$\mu$ of degree $-2$ such that $d\mu =\gam$. Then $\mu |_{K^2}$ is
a cocycle and one can find (as in 4.2) a Hochschild
 2-cocycle $\mu '$ of degree
$-2$ whose resriction to $K^2$ is equal to $\mu$. Now it is easy
to see that if we define $f_2$ to be equal to $\tilf _2-\mu +\mu'$,
then it will satisfy the above identity.

\subheading{4.5. Extension to the third level}
Let $f_2$  be the second term
of the deformation.
The second obstruction is given by
the cocycle
$$
\del =f_1\circ (f_2\otimes id-id\otimes f_2)-
f_2\circ (f_1\otimes id-id\otimes f_1).
$$
One can easily verify that (iii) is equivalent to the
vanishing of $\del$ in $H^3(A,A)$. Hence,
we can continue our deformation to the third level.

\subheading{4.6} Now we see that Proposition 3.7 implies that our deformation
extends to some (unique)
graded deformation of $A$.

Let us now prove that the fiber $\tilU$ of this deformation at $t=1$
is isomorphic to $U$. We have a
map of vector spaces $\tilvarphi$: $V\to\tilU$, which by the
conditions imposed on $f_1$ and $f_2$ gives rise to a map $\varphi$:
$U\to \tilU$ , since the multiplication in $A_t$ is represented as
$$
a\times b=ab+\sum^\infty_{i=1} f_i(a,b)t^i.
$$
We know that there is a canonical isomorphism  between $A$ and gr$(\tilU)$.
Obviously, the composition map $$A\buildrel p\over\to grU \buildrel
gr\varphi\over\to gr\tilU\to A$$  is $id_A$.  Thus $p$,  being a surjection,
is an isomorphism and $\varphi$ is an isomorphism, too. $\bigsquare$

\subheading{4.7. Remark} In fact one can prove that the deformation
that we have constructed is actually isomorphic to the Rees ring of $U$
(see 1.1.)

\sect{5. Examples}

\subheading{5.1. The classical PBW-theorem} Let
$A=S(V)$. Then $R=\Lam^2(V)$. Consider the case $\bet \equiv 0$. Then a
 straightforward computation
shows that heading (ii) of Lemma 3.3 is the Jacobi identity for $\alp$.
Thus, $V$ with the bracket $\alp$ becomes a Lie algebra
and so 4.1 gives us another proof of the classical PBW theorem.

\subheading{5.2. The supercase} Let $V=V_{\overline 0}\oplus V_{\overline
1}$ be a
superspace and $A:=S(V_{\overline 0})\otimes \Lam (V_{\overline 1})$ be its
symmetric algebra.
Suppose as in 5.1 that $\bet \equiv 0$. Then
condition (ii) is the super-Jacobi identity for $\alp$.
Thus, we see that $\alp$ defines a Lie superalgebra structure
on $V$ (note that conditions (i) and (ii) provide us with a definition of a Lie
superalgebra over a field of any characteristic); hence, 4.1 proves also the
superversion of the PBW theorem (the fact that $A$ is a Koszul algebra follows,
for example, from Appendix).

\subheading{5.3. Weyl and Clifford algebras}
Let $A$  be either $S(V)$ or $\Lam (V)$ and let $\alp$  be
identically zero.  Then the only restriction on $\bet$ is $\bet\otimes
id-id\otimes\bet =0$ that  any map $\bet: R\to k$ clearly satisfies (in our
case $R$ is either $S^2(V)$ or $\Lam ^2(V)$.
Hence, Theorem 4.1 proves also the PBW theorem for Weyl and Clifford algebras.


\subheading{5.4. The standard complex}
We conclude this section with a generalization of the standard complex for Lie
algebras to quadratic algebras of PBW-type.
Let $U=Q(V,P)$ be of PBW type and the corresponding homogeneous quadratic
algebra be Koszul. Suppose that $P\subset T^2\oplus T^1$.
Then there exists a Kozsul complex  $\tilK^i(U)$ which gives a free
resolution of $U$ in the category of $U$-bimodules.
Let us define $\tilK^i$ to be $U\otimes K^i\otimes U$, where
the $K^i$ are as in 3.6  for the corresponding homogeneous
quadratic algebra. The condition $(\alpha\otimes id-id\otimes\alpha)=0$
implies that the differential of the bar-resolution of $U$
maps $\tilK^i$ into $\tilK^{i-1}$.

To prove that the differential is exact, it suffices to notice that
gr $(\tilK^\cdot)$ is canonically isomorphic to the Koszul complex of the
corresponding homogeneous quadratic algebra. Since passage to the associated
graded module as a functor is faithfull, the assertion follows.

As a corollary of this fact, we obtain, by tensoring with $k$,
a free resolution of $k$ in the category of left $U$-modules.
For example, in the case when $U=U(\grg)$ (the universal enveloping algebra
of $\grg$, where $\grg$ is a Lie algebra, we obtain the standard complex of
$\grg$.
\sect{Appendix}

Here we present a brief review of some basic properties
 of Koszul algebras.

\subheading{A.1} Let $A$ be a homogeneous quadratic algebra over a field $k$.
 We start with the following (traditional) definition of Koszul algebras.
 (In what follows we will prove that it is equivalent to the definitions given
 in 3.4 and 3.6.) Let us view $k$ as  a left $A$-module.

\subheading{Definition}A quadratic algebra $A$ is called {\it Koszul} if
$Ext^i_j(k,k)=0$ for $i\not=-j$ (where $Ext^i(k,k)$ is taken in the cateogry
of left $A$-modules).

\subheading{A.2. Proposition}
{\sl Let $\tilK^\cdot(A)$ denote the Koszul complex introduced in \S  3.
The following conditions are equivalent:

a) $A$ is a Koszul algebra (in the sense of A.1).

b) $\tilK^\cdot$ is a resolution of $A$ in the category of $A$-bimodules.

c) $\tilK^{\cdot}\otimes_A k$ is a resolution of $k$ in the category of left
$A$-modules.

d) $H^i_j(A,M)$, where $A$ is considered as an $A$-bimodule, vanishes for
any $\ZZ ^+$-graded $A$-bimodule $M$ and $i<-j$.

e) $Ext^i_j(k,N)$, where $k$ is considered as a left $A$-module,
vanishes for any $\ZZ ^+$-graded $A$-module $N$ and $i<-j$.}

\subheading{Remark} If one of the equivalent conditions of A.2
is satisfied, then $\tilK^{\cdot}\otimes_A k$ is the Koszul resolution of
$k$ in the sense of [Pr].

\subheading{A.3} In the proof of the above proposition we will
need the following general result.

\subheading{Lemma}. {\sl Let $P^\cdot =\oplus
_{i=-\infty}^{\infty}P_i^{\cdot}$ be a
graded complex of graded $A$-modules over a $\ZZ ^+$-graded $k$-algebra $A$
with
$A_0=k$. Suppose that these graded $A$-modules are free with homogeneous
generators and
with homogeneous of degree 0 differentials and such that $P^{\cdot}_i=0$
for $i<<0$. Suppose that $\tilP^{\cdot}=P^{\cdot}\otimes_A k$ is exact. Then
 $P^{\cdot}$ is exact.}
\medskip

\noindent {\bf Proof}.
Let us denote by $P^\cdot_i$ the subcomplex of $P^\cdot$,
whose chain groups $P^j_i$ are the subgroups of $P^j$
generated by elements of degree $\le i$.
Then $P^\cdot$ is the direct limit of $P^\cdot_i$, so it will
be enough to prove that if
$\tilP^\cdot_i$=$P^\cdot_i\otimes_A k$ are exact, then all $P^\cdot_i$ are
exact
because they $P^{\cdot}_i$ is a direct summand of $P^\cdot\otimes_Ak$.
So, we are reduced to the case when the chain
groups in $P^\cdot$ are generated by elements of degree $\le i$.

Let us
fix this $i$ and proceed by the descending induction on $j$:
$j$ is the minimal $m$ for which $P^\cdot$ is nontrivial in
degree $m$. (This $j$ is well-defined since $P^{\cdot}_i$ is
assumed to be zero for $i<<0$.)

  For $j=i+1$ the assertion is trivially satisfied, so we
  have to carry out the inductive step.

Let the assertion be true for $j+1$; let us prove that
it is true for $j$. Let $P^\cdot_j$ be as before. Then $P^\cdot/P^\cdot_j$
satisfies the inductive hypothesis, and $P^\cdot_j$ is exact since it is
equal to $\tilP^\cdot_j\otimes_kA$, by the minimality of $j$.
Hence, the assertion follows. $\bigsquare$

\subheading{A.4. Proof of Proposition A.2}
  (e)$\Rightarrow$(d): follows from the fact that $H^i_j(A,M)$ (in the
  category of $A$-bimodules) is naturally isomorphic to
  $Ext^i_j(k,M)$ (in the category of left $A$-modules)
  for every bimodule $M$.

  Implications (b)$\Rightarrow$(d),
  (c)$\Rightarrow$(e)
  follow immediately from exactness of the Koszul complex.

  (b)$\Rightarrow$(c) is easy, since we can tensor the
  Koszul resolution of $A$ as an $A$-bimodule by $k$ and
  obtain the Koszul resolution of $k$, as a left $A$-module.

  The implication (c)$\Rightarrow$(b) is a direct corollary of Lemma A.3.

  Implication (c)$\Rightarrow$(a) is a consequence of the
  exactness of the Koszul complex.

  (a)$\Rightarrow$(e):
  Let $P^\cdot$ be a minimal free graded resolution of $k$. Then one has
  $Ext^i(k,k)={\rm Hom}_A(P^i,k)$.
In particular, $Ext^i_j(k,k)=0$ for $i<-j$ imply that $P^i$ is trivial in
degrees
$i<j$, so $Ext^i_j(k,N)$ vanishes for every module $N$ when $i<-j$.

In order to prove (e)$\Rightarrow$(c) we shall proceed by induction: let
$$
A\otimes K^i\longrightarrow A\otimes K^{i-1}
\longrightarrow\cdots\longrightarrow A\otimes V\longrightarrow k
$$
be exact. Then we see that $Ext^{i+1}_j(k,N)=0$ for any $N$ and also $j>i+1$
implies that $\ker \left(A\otimes\tilK^i\longrightarrow A\otimes\tilK^{i-1}
\right)$ is generated by the elements of degree $\le i+1$, i.e.,  these
generators lie in $V\otimes K^i$$\subset$$V^{\otimes{i+1}}$.

At the same time, they have to lie in $R\otimes{V^{\otimes{i-1}}}$.
In other words, the generators of this kernel lie in
$R\otimes K^{i-2}$$\cap$$V\otimes K^{i-1}$, so $A\otimes K^{i+1}$
is surjectively mapped onto the kernel. This finishes the proof
of Proposition. $\bigsquare$

\subheading{A.5. The dual (co)algebra} We conclude this appendix with a
description
of some properties of {\it the dual algebra of a quadratic
algebra} (see [Ma]). Let us start with a homogeneous quadratic algebra
$Q(V,R)$.
Then we may consider a graded subspace of $T(V)$ , namely $\oplus_{i=0}
^{\infty} K^i$.
This subspace admits a natural coalgebra structure,
induced from that of $T(V)$.

Suppose now that $V$ is finite dimensional. Then we may  also
consider a dual object to this coalgebra, namely $\oplus _{i=0}
^{\infty}K^{i*}$. It has a natural algebra structure and it is
called the {\it dual quadratic algebra to} $A$ and denoted by $A^!$.

\subheading{A.6. Proposition} {\sl $A$ is a Koszul
algebra if and only if $A^!$  is; in this case $Ext^i_j(k,k)=\cases{0&for
$i\not= -j$\cr
A^!_i&for $i=j$}$.}
\medskip

\noindent {\bf Proof}. It clearly suffices to prove the statement in one
direction, since $A=(A^!)^!$. Consider the Kozsul complex for $A$ in degree
$i$:
$$
(A^!_i)^*\longrightarrow\cdots\longrightarrow
(A^!_{i-k})^*\otimes{A_k}\longrightarrow\cdots \longrightarrow A_i.
$$
It is exact by the assumption. When we dualize it with respect
to $k$, we obtain the Kozsul comlex for $A^!$ in degree $i$. $\bigsquare$

\medskip
\sect{References}
\smallskip

[G] M.Gerstenhaber, Deformations of rings and algebras,
{\it Ann. of Math.(2)}, {\bf 79}, 1964, 59-103
\smallskip

[GeS] M.Gestenhaber and S.D.Schack, Algebraic cohomology and
deformation theory, in: "Deformation theory of algebras and
structures and applications",
M.Hazevinkel and M.Gestenhaber eds., Kluwer, Dodrecht, 1988, 11-264.
\smallskip

 [Ma] Y.I.Manin, Quantum groups and noncommutative geometry, Montreal
University, 1988.
\smallskip

[PoP] A.E.Polishchuk and L.E.Positselsky, On quadratic algebras
(in Russian), preprint, Moskow University, 1990.
\smallskip

 [Pr] S.Priddy, Koszul resolutions, {\it Trans. AMS}, {\bf152}, 1970, 39-60

\end